\documentclass[aps,prl,twocolumn,superscriptaddress,amsmath,amssymb]{revtex4-1}

\usepackage[latin1]{inputenc}
\usepackage{graphicx}
\usepackage{dcolumn}
\usepackage{bm}
\usepackage{color}
\usepackage{tabularx}
\newcolumntype{Y}{>{\centering}X}
\usepackage[cdot,textstyle,amssymb]{SIunits}

\usepackage{sistyle}
\SIunitspace{\cdot}
\SIunitdot{\cdot}
\SIdecimalsign{.}
\usepackage{ulem}

\usepackage[pdfborder={0 0 0}, colorlinks=true,citecolor=blue,linkcolor=black,urlcolor=blue]{hyperref}

 \definecolor{orange}{rgb}{1, 0.625, 0.0625}

\begin{document}


\title{High-field thermal transport properties of the Kitaev quantum magnet $\alpha$-RuCl$_3$: evidence for low-energy excitations beyond the critical field}





\author{Richard Hentrich}
\affiliation{Leibniz Institute for Solid State and Materials Research, 01069 Dresden, Germany}

\author{Xiaochen Hong}
\affiliation{Leibniz Institute for Solid State and Materials Research, 01069 Dresden, Germany}

\author{Matthias Gillig}
\affiliation{Leibniz Institute for Solid State and Materials Research, 01069 Dresden, Germany}

\author{Federico Caglieris}
\affiliation{Leibniz Institute for Solid State and Materials Research, 01069 Dresden, Germany}

\author{Matija \v{C}ulo}
\affiliation{High Field Magnet Laboratory (HFML-EFML), Radboud University Nijmegen, 6525 ED Nijmegen, The Netherlands}

\author{Maryam Shahrokhvand}
\affiliation{High Field Magnet Laboratory (HFML-EFML), Radboud University Nijmegen, 6525 ED Nijmegen, The Netherlands}

\author{Uli Zeitler}
\affiliation{High Field Magnet Laboratory (HFML-EFML), Radboud University Nijmegen, 6525 ED Nijmegen, The Netherlands}

\author{Maria Roslova}
\affiliation{Faculty of Chemistry and Food Chemistry, TU Dresden, 01062 Dresden, Germany}
\author{Anna Isaeva}
\affiliation{Leibniz Institute for Solid State and Materials Research, 01069 Dresden, Germany}
\affiliation{Faculty of Chemistry and Food Chemistry, TU Dresden, 01062 Dresden, Germany}
\author{Thomas Doert}
\affiliation{Faculty of Chemistry and Food Chemistry, TU Dresden, 01062 Dresden, Germany}

\author{Lukas Janssen}
\affiliation{Institut f\"ur Theoretische Physik, TU Dresden, 01062 Dresden, Germany}

\author{Matthias Vojta}
\affiliation{Institut f\"ur Theoretische Physik, TU Dresden, 01062 Dresden, Germany}

\author{Bernd B\"uchner }
\affiliation{Leibniz Institute for Solid State and Materials Research, 01069 Dresden, Germany}
\affiliation{Institute of Solid State Physics, TU Dresden, 01069 Dresden, Germany}
\affiliation{Center for Transport and Devices, TU Dresden, 01069 Dresden, Germany}

\author{Christian Hess}\email{c.hess@uni-wuppertal.de}
\affiliation{Leibniz Institute for Solid State and Materials Research, 01069 Dresden, Germany}
\affiliation{Center for Transport and Devices, TU Dresden, 01069 Dresden, Germany}
\affiliation{Fakult\"at f\"ur Mathematik und Naturwissenschaften, Bergische Universit\"at Wuppertal, 42097 Wuppertal, Germany}


\date{\today}


\begin{abstract}
We investigate the phononic in-plane longitudinal low-temperature thermal conductivity $\kappa_{ab}$ of the Kitaev quantum magnet $\alpha$-RuCl$_3$ for large in-plane magnetic fields up to 33~T.
Our data reveal for fields larger than the critical field $B_c\approx8$~T, at which the magnetic order is suppressed, a dramatic increase of $\kappa_{ab}$ at all temperatures investigated. The analysis of our data shows that the phonons are not only strongly scattered by a magnetic mode at relatively large energy which scales roughly linearly with the magnetic field, but also by a small-energy mode which emerges near $B_c$ with a square-root-like field dependence.
While the former is in striking agreement with recent spin wave theory (SWT) results of the magnetic excitation spectrum at the $\Gamma$ point, the energy of the latter is too small to be compatible with the SWT-expected magnon gap at the $M$ point, despite the matching field dependence. Therefore, an alternative scenario based on phonon scattering off the thermal excitation of random-singlet states is proposed.
\end{abstract}


\pacs{}

\maketitle

\section{Introduction}

For more than a decade, the celebrated Kitaev model describing spin-1/2 degrees of freedom on a honeycomb lattice with bond-dependent interactions \cite{Kitaev2006} has been attracting much attention. This strongly frustrated spin model is exactly solvable and has been shown to exhibit a gapless quantum spin liquid (QSL) ground state with spin excitations which fractionalize into localized $Z_2$ gauge fluxes and itinerant Majorana fermions \cite{Kitaev2006,Baskaran2007,Knolle2014}. An external magnetic field renders the ground state a topological quantum spin liquid (TQSL), with peculiar properties. Chiral Majorana edge modes arise within the field-induced gap, and the $Z_2$ vortices acquire non-Abelian anyonic statistics, rendering Kitaev materials relevant for quantum computing \cite{Kitaev2006}.

Substantial effort has been devoted towards experimentally realizing Kitaev's spin liquid. Candidate materials are Ir and Ru compounds where considerable spin-orbit interaction leads to the emergence of Kitaev interaction between spin-orbit entangled $j_\mathrm{eff}=1/2$ moments \cite{Jackeli2009,Singh2010,Takagi2019}. However, most of these compounds exhibit a magnetically ordered ground state due to the presence of additional interactions in the Hamiltonian \cite{Chaloupka2010,Gotfryd2017,Winter2017a}.
Consequently, ways of suppressing this order in favor of a spin liquid have been sought for, and
one of the most promising materials here is $\alpha$-RuCl$_3$. It displays zigzag antiferromagnetic order in zero magnetic field below $T_N\approx7$~K \cite{Kubota2015,Sears2015,Cao2016}. This long-range order can be suppressed by an in-plane field of $B_c\approx 8$~T, where signatures of quantum criticality have been reported \cite{Wolter2017,Sears2017,Zheng2017,Lampen-Kelley2018}. At fields larger than $B_c$, magnetic order remains absent, and it has been speculated that a field-induced TQSL exists for $B>B_c$ in a window of magnetic fields \cite{Wolter2017,Baek2017,Banerjee2018}. This notion seems to be spectacularly corroborated by the recent report of half-integer quantization of the thermal Hall conductance near $B_c$ \cite{Kasahara2018a}.
Independently, there is evidence that the magnetic excitation spectrum at $B>B_c$ is adiabatically connected to the polarized state realized in the high-field limit: Field-dependent magnon-type excitations at the $\Gamma$ point, as observed by electron spin resonance (ESR) \cite{Ponomaryov2017,Wellm2018} and Raman spectroscopy \cite{Sahasrabudhe2020}, as well as a field-dependent low-energy gap extracted from specific-heat data \cite{Wolter2017} appear consistent with semi-classical spin-wave theory (SWT) and exact diagonalization (ED) results \cite{Wolter2017,Winter2018,Janssen2019} for extended Heisenberg-Kitaev-Gamma models. Here, the high-field magnon spectrum is characterized by a dispersion minimum at the $M$ point and a van Hove singularity at the $\Gamma$ point.

Magnetic excitations affect the heat transport in solids in two distinct ways: They carry heat themselves, leading to a magnetic contribution to thermal conductivity which may expose important information about the excitations' heat capacity and scattering properties \cite{Hess2019,Kolland2012,Toews2013}. In addition, magnetic excitations act as scattering channels for other heat-carrying modes, most importantly phonons. If thermal conductivity is dominated by the latter,  then thermal conductivity carries information about spin-phonon scattering \cite{Hofmann2001,Jeon2016}.

For $\alpha$-RuCl$_3$, previous studies have revealed that the longitudinal heat conductivity is primarily phononic, with substantial spin-phonon coupling leading to strongly field- and temperature-dependent spin-scattering of the phonons \cite{Hentrich2018,Yu2018}. Considerable controversy exists, however, with respect to a sizeable transversal heat conductivity that has been observed in the material for out-of-plane magnetic fields \cite{Kasahara2018,Hentrich2019}. On one hand, this has been interpreted as evidence for thermal transport by Majorana fermions \cite{Kasahara2018}, which directly connects to the afore-mentioned half-integer quantization of the thermal Hall conductance \cite{Kasahara2018a}. On the other hand, it has been pointed out that a purely phononic origin cannot be excluded \cite{Hentrich2019,Li2020}. In such a case, several theories for the phonon thermal Hall effect invoke a particular role of spin-phonon interaction \cite{Sheng2006,Kagan2008,Mori2014}.

Motivated by these results, we investigate in this work the in-plane longitudinal thermal conductivity $\kappa_{ab}$ for in-plane magnetic fields in an extended magnetic field and temperature range, namely fields up to 33~T, from sub-Kelvin temperatures up to 80~K.
Our data reveal that for $B>B_c$, the heat conductivity dramatically increases with the magnetic field at all temperatures investigated. These results provide strong evidence that the phonons are scattered not only by a magnetic mode at the $\Gamma$ point,  but additionally by further, quite different magnetic excitations. While the energy of the former, as already inferred previously \cite{Hentrich2018}, strongly increases with increasing field above $B_c$, reaching about $100$\,K at $30$\,T, the energy of the latter is more than an order of magnitude smaller. As we show, this appears inconsistent with spin-wave modes of the high-field phase. We conjecture that this scattering arises from a small concentration of residual defects which create random-singlet-type excitations below the bulk energy gap of the high-field phase, possibly consistent with NMR results showing persistent relaxation at temperatures below the bulk gap \cite{Baek2017,Baek2020}.

\section{Experimental Details}

Single crystals of $\alpha$-RuCl$_3$ have been grown by chemical vapor transport \cite{Hentrich2018}.
For 8~K $\leq T \leq $ 80~K,  $\kappa_\mathrm{ab}$-measurements were performed in a home-built probe, employing a 4-points measurement geometry.
One face of the rectangularly shaped single crystal (sample~1) was thermally excited by a resistive chip heater, and the resulting temperature gradient across the sample picked up by a differential Au/Fe-Chromel thermocouple. The field calibration of the thermocouple was obtained in-situ by monitoring the temperature gradient of a glass sample (Herasil\textsuperscript{\textregistered}) in an identical 4-points configuration with a second thermocouple of identical build, assuming a field independent thermal conductivity for this nonmagnetic material. Data were taken up to 30~T at constant temperatures with the magnetic field ramped in 2~T steps, with the required holding time to allow for thermal equilibrium to be reached at each $B$- and $T$-value.

Measurements at $T < 4$~K were performed on a second single crystal (sample~2). A closed-cycle $^3$He-cooled setup was used, employing a standard steady-state one-heater, two-sensors (Cernox\textsuperscript{\textregistered} chip thermometers) geometry. The two thermometers were calibrated in-situ, where data were taken up to 33~T in an analogous manner to the high-$T$ ($T \gtrsim 8$~K) measurements. For both samples, magnetic fields were applied in-plane, directed perpendicular to the heat current.
All high-field measurements have been conducted at HFML. For sample~1, additional measurements with $B\leq16$~T have been performed at IFW Dresden using the same setup, yielding field dependent data in excellent agreement with the HFML results for the overlapping field range. The three data sets obtained show good consistency and are presented in this work modulo a fixed scaling factor for direct comparability.

\section{Results}

\begin{figure}[t]
\centering
\includegraphics[width=\columnwidth]{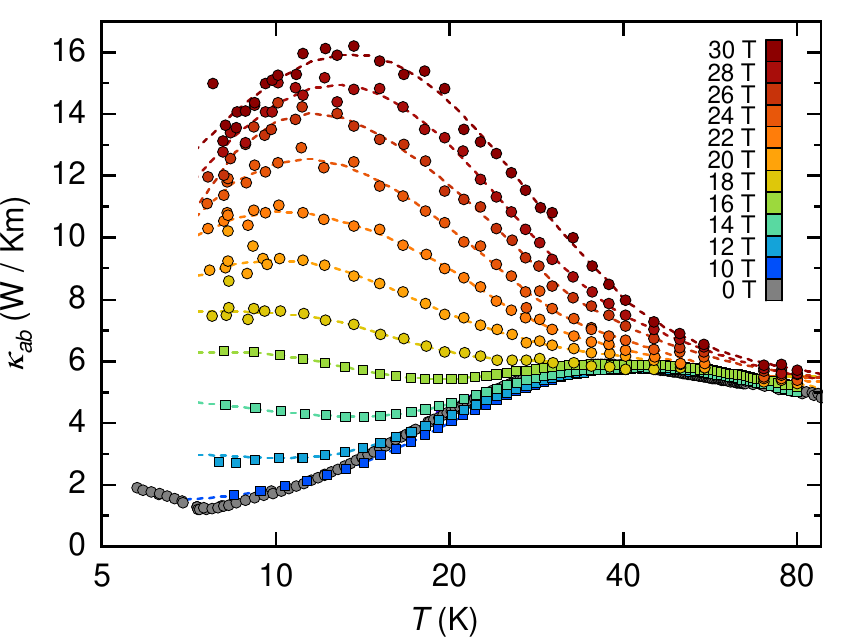}
\caption{
Temperature dependence of the longitudinal heat conductivity $\kappa_{ab}$ of $\alpha$-RuCl$_3$ parallel to the honeycomb layers in zero and high in-plane magnetic fields.
Data for 10~T~$\le B \le$~16~T were taken at IFW Dresden (squares), data for the same sample, $B=0$~T and $B \ge 18$~T at HFML (circles). The temperature scale is logarithmic, dashed lines are guides to the eye.}
\label{fig:kappavst}
\end{figure}

\begin{figure*}[t]
\centering
\includegraphics[width=\textwidth]{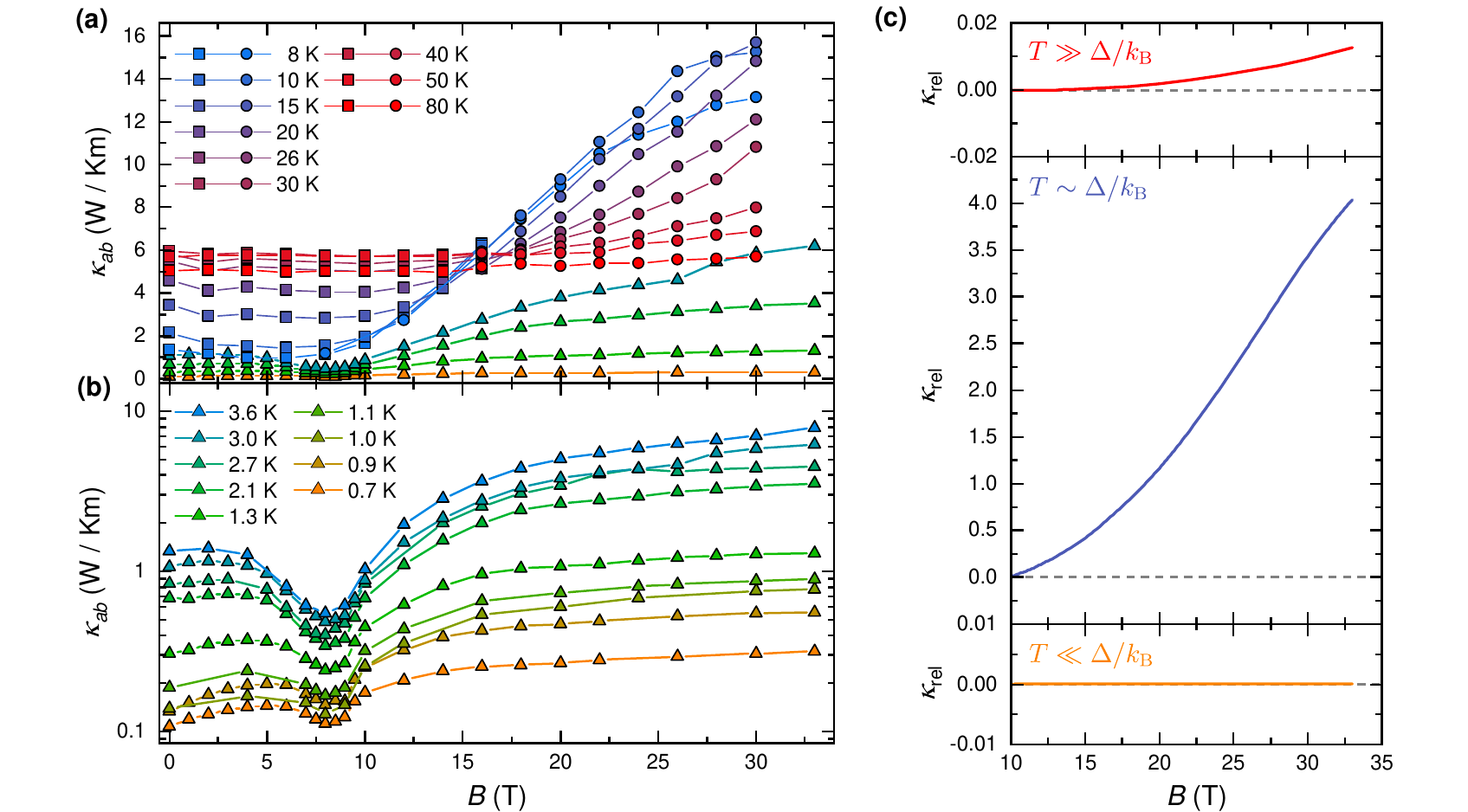}
\caption{(a) and (b) Magnetic field dependent data of the longitudinal heat conductivity $\kappa_{ab}$ of $\alpha$-RuCl$_3$ parallel to the honeycomb layers at constant $T$. Data of sample 1 were taken at IFW Dresden (open squares) as well as at HFML (open circles), sample 2 was measured at HFML (open triangles).  (a) At $B\gtrsim B_c$, $\kappa_{ab}$ increases strongly at all temperatures measured. (b) Semi-logarithmic plot of the low-$T$ data - a substantial field dependence up to highest field is visible down to the lowest temperatures, as discussed in the main text. (c) Illustration of the expected relative field dependencies $\kappa_\mathrm{rel}=(\kappa_{ab}(B)-\kappa_{ab}(10\mathrm{T}))/\kappa_{ab}(10\mathrm{T})$ at three distinct temperature regimes (see text) for a model comprising one characteristic magnetic scattering mode of energy $\Delta$. Details on the calculations are specified in the Appendix.}
\label{fig:kappavsb}
\end{figure*}

We first address the in-plane thermal conductivity $\kappa _\mathrm{ab}$ for temperatures $T\gtrsim 5$~K in high external magnetic in-plane fields 0~T~$\le B \le$~30~T as shown in Fig.~\ref{fig:kappavst}. For $B\leq18$~T these data are very similar to our previous results for the same temperature range \cite{Hentrich2018}:
At zero magnetic field (gray circles),  $\kappa _\mathrm{ab}$ exhibits first a sharp kink at the magnetic ordering temperature at about 7.7~K, followed by a broad peak at about 40~K.
Applying an in-plane external magnetic field $B\gtrsim B_c$ causes a strong enhancement of $\kappa _\mathrm{ab}$, with the increase below $T \approx 40$~K being most prominent. Remarkably, the further enhancement of the magnetic field up to 30~T causes a mere continuation of the field enhancement of $\kappa _\mathrm{ab}$ lacking any signs of saturation, resulting in a fivefold amplification of $\kappa _\mathrm{ab}$ at $T=15$~K as compared to its zero field value (see Fig.~\ref{fig:kappavst}).

\begin{figure*}[thb]
\centering
\includegraphics[width=0.75\textwidth]{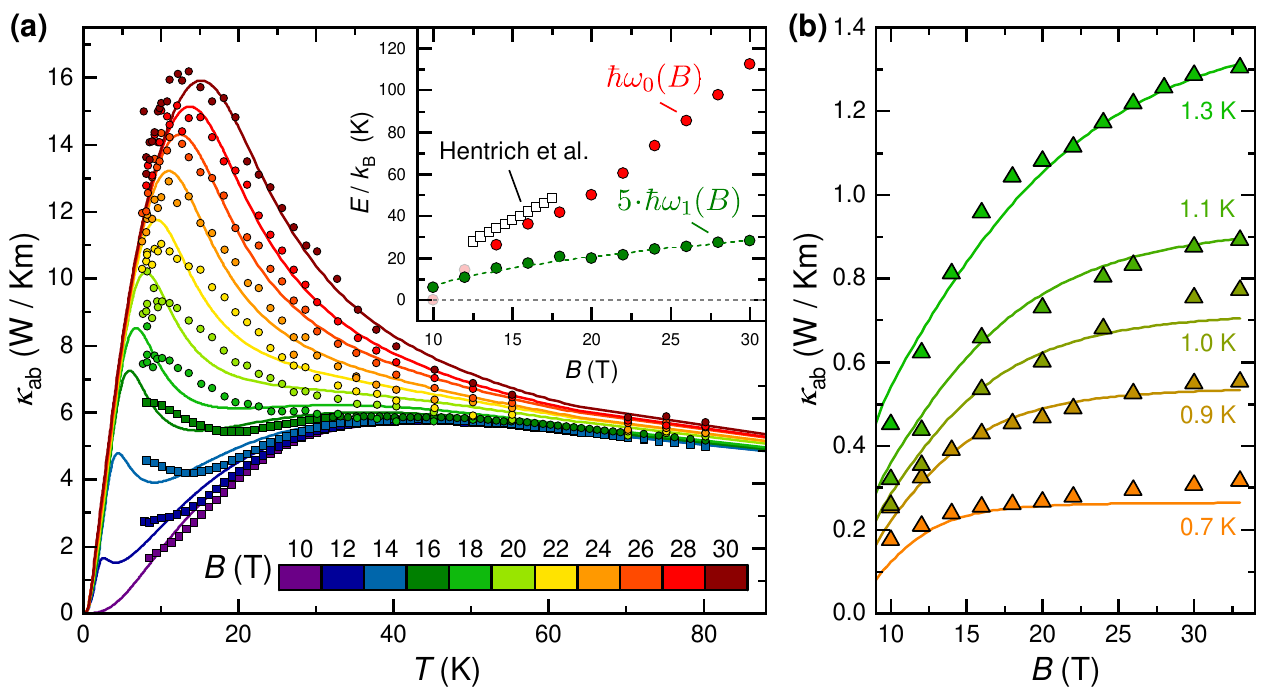}
\caption{ (a)  Temperature dependence of $\kappa_{ab}$ of $\alpha$-RuCl$_3$ (open symbols, sample 1) for $T \gtrsim 8$~K with fits to the two-mode model. The inset shows the field dependence of the obtained energy scales $\hbar \omega _0 (B)$ (red) and $\hbar \omega_1 (B)$ (green, scaled  for better visibility). The square root fit to the low-energy excitation's energy is indicated by the dashed line. Open squares are literature data for $\hbar \omega _0 (B)$, reproduced from Ref.~\cite{Hentrich2018}. (b) Low-temperature magnetic field dependence of $\kappa_{ab} $ (open symbols, sample 2), lines are fits as described in the text. }
\label{fig:kappafits}
\end{figure*}

The non-saturating field dependence  $\kappa _{ab}(B)$ at fixed temperature can be observed more accurately in Fig.~\ref{fig:kappavsb}~(a).
While a small decrease of  $\kappa _{ab}(B)$ is discernible at $B\lesssim B_c$, larger fields $B\gtrsim B_c$ cause a strong increase of $\kappa _{ab}(B)$ for all $T \ge 8$~K. The slope $ \partial \kappa _{ab} / \partial B$ is highest at 8 to 15~K, then gradually decreases with increasing $T$ where even at $80$~K a finite slope still is detectable.

Qualitatively, the observed $\kappa _{ab}(T,B)$ is consistent with an all-phononic heat conductivity which is affected by a strong, $B$- and $T$-dependent spin-phonon scattering. More specifically, as has been shown in Ref.~\cite{Hentrich2018}, at fields above about 8~T a single magnetic mode, the energy $\hbar\omega_0$ of which increases about linearly with the field, is capable of causing a concomitant strong enhancement of $\kappa _{ab}(T,B)$ if the heat transporting phonons of the mode's energy are strongly scattered off it. As mentioned in the introduction, the existence of such a magnetic mode has been confirmed by several probes including NMR \cite{Baek2017}, ESR \cite{Ponomaryov2017,Wellm2018}, Raman \cite{Sahasrabudhe2020} as well as neutron scattering \cite{Balz2019}, and is also predicted theoretically in SWT and ED model calculations, where it is connected to a van Hove singularity near the $\Gamma$ point \cite{Winter2018,Wolter2017,Janssen2019}. The non-saturated
$\kappa _{ab}(B)$ even at 30~T therefore suggests that the energy of this mode continues to increase linearly beyond 18~T, up to the highest field measured.


In such a scenario, the reduction of the slope $ \partial \kappa _{ab}/ \partial B$ with increasing temperature naturally results from the scattered phonons' reduced contribution to the total phonon heat conductivity: This is because the phonons predominantly carrying the heat have an energy $\hbar\omega\sim k_\mathrm{B}T$  \cite{Callaway1959,Callaway1961,Hentrich2018}.
Thus, for $T\rightarrow\infty$, the field dependence of $\hbar\omega_0$ becomes increasingly unimportant for $\kappa _{ab}(T,B)$ and leads to a vanishing slope as soon as $T > \hbar\omega_0(B) / k_\mathrm{B}$.
To illustrate this temperature effect, Fig.~\ref{fig:kappavsb} shows the heat conductivity's expected relative magnetic field dependence $\kappa_\mathrm{rel}$ for three temperature regimes,
with a single magnetic scattering mode of energy $\Delta$ present in the system. Consequently, for the case $T \gg \Delta / k_\mathrm{B}$ shown in the top panel, the strong field increase present at intermediate temperatures ($T \sim \Delta / k_\mathrm{B}$, center panel) is reduced to less than 2~\% over the investigated field range. The corresponding limit is reached in the experimental data approximately at 80~K, see Fig.~\ref{fig:kappavsb}~(a).
In a similar fashion to the high temperature case one can argue that, towards $T \rightarrow 0$, one would expect a rapidly diminishing slope $ \partial \kappa _{ab} / \partial B$ (see bottom panel of Fig.~\ref{fig:kappavsb}~(c)).

One can test the latter expectation quite conveniently by probing the thermal conductivity at about 1~K. Here, $T\ll\hbar\omega_0(B) / k_\mathrm{B}$ should be always fulfilled since $\hbar\omega_0(B)/k_\mathrm{B}\gtrsim 30$~K at $B>12$~T \cite{Hentrich2018,Ponomaryov2017}. Figs.~\ref{fig:kappavsb}~(a) and (b) show our pertinent low-temperature measurements (sample~2) for $T<4$~K as a function of the magnetic field up to 33~T. While the data in the magnetically ordered phase at $B\lesssim 7.5$~T show a similar non-monotonic field dependence as reported previously \cite{Yu2018,Lampen-Kelley2018}, for a magnetic field beyond $\sim8$~T, $\kappa _{ab}(B)$ still substantially increases with the magnetic field. This entails a sizeable positive slope of $\kappa _{ab}(B)$ even at $0.7$~K and high $B>20$~T, in clear contradiction to the expectation described above and depicted in the lower panel of Fig.~\ref{fig:kappavsb}~(c).

Clearly, this main finding of our work, i.e., the unexpected field-induced enhancement of the low-temperature thermal conductivity, provides new information about the low-energy excitations of $\alpha$-RuCl$_3$. A new transport channel carried by magnetic excitations rising in field appears unlikely, since the increasing Zeeman energy naturally depletes magnetic low-energy excitations in the quantum disordered phase at $B\gtrsim B_c$. Thus, the only way to rationalize the observed enhancement is to understand it similarly to the higher-temperature findings as a result of a field-induced reduction of the phonon scattering. However, since the known lower bound  $\hbar\omega_0(B)$ of magnetic excitation energies of $\alpha$-RuCl$_3$ is far too large as to cause the observed field dependence, we conjecture that additional magnetic scattering at energies $\hbar\omega_1$ persists with $\omega_1\ll\omega_0$, even at 33~T.

In our previous work, we had employed a modified Callaway model for describing the in-plane field dependence of $\kappa _{ab}$ by introducing a magnetic scattering term to the conventional expressions for phonon-phonon, phonon-defect and phonon-boundary scattering \cite{Hentrich2018}.
More specifically, we had used an empirical field-dependent magnetic phonon scattering rate
\begin{equation}
 \tau_\mathrm{mag,0}^{-1}= C_0  \Theta(K-\hbar\omega)\frac{\exp\left({-\frac{\hbar\omega_0  }{k_\mathrm{B}T}}\right)}{1+3\exp\left({-\frac{\hbar\omega_0}{k_\mathrm{B}T}}\right)}
\label{eq:triplet}
\end{equation}
which mimics phonons scattering off a magnetic excitation continuum effective in zero field in the energy range $0\leq \hbar\omega \leq |K|$. Here, $K$ is the Kitaev interaction, $\Theta$ the step function, and $C_0$ a field-dependent parameter measuring the scattering strength. The energy scale $\hbar\omega_0$ plays the role of a field-induced low-energy cut-off.
Indeed, using the high-temperature ($T \gtrsim 8$~K), high-field data, good fits can be obtained with parameters consistent with our previous results (see Fig.~\ref{fig:kappafits}~(a), fit parameters are presented in the Appendix. Remarkably, as to be expected from our qualitative discussion of the high-field data, our fit yields $\omega_0(B)$ to increase approximately linearly up to 30~T (inset of Fig.~\ref{fig:kappafits}~a).

To account for additional low-energy scattering in the model, as explained above, a straightforward approach is to introduce a second magnetic scattering rate $\tau_\mathrm{mag,1}^{-1}$ of identical form as Eq.~\eqref{eq:triplet}, with the coupling $C_1$ and the field-dependent energy scale $\hbar\omega_1$, such that with
\begin{equation}
\tau_\mathrm{mag}^{-1}=\tau_\mathrm{mag,0}^{-1}+\tau_\mathrm{mag,1}^{-1}~,
\label{eq:tausum}
\end{equation}
a two-mode model is defined.

Because the two magnetic scattering modes contribute to the field dependency of $\kappa_{ab}$ at well separated temperature regimes, the very low temperature data ($T < 1.5$~K) can be fit by the two-mode model with $\tau_\mathrm{mag,0}^{-1}$ set to zero, as the corresponding scattering becomes significant only at higher energies/temperatures. This leaves only a single magnetic scattering term for fitting multiple $\kappa _{ab}(B)$ curves while maintaining the previously obtained phononic parameters. As shown in Fig.~\ref{fig:kappafits}~(b), it is indeed possible to reproduce the low-temperature field dependence. With this fitting approach, we find that $\hbar\omega_1(B)/k_\mathrm{B}$ varies in a sublinear fashion with $B-B_c$ and reaches approximately 6~K at 33~T (see Appendix for the fit parameters and a comparison of various power law fits). Moreover, the scattering intensity of the low-energy excitation, as quantified by the prefactor $C_1$, appears to be by a factor of $50$ smaller than that of the higher-energy scattering mechanism. This is also the reason why including the second magnetic scattering term in the modelling does not change the conclusions concerning the dominant high-energy scattering mechanism.

\section{Discussion and summary}

As already mentioned above, the large energy scale $\hbar\omega_0$ with its linear field dependence is in almost perfect agreement with magnon-type excitations at the $\Gamma$ point, as probed by ESR and Raman scattering \cite{Ponomaryov2017,Wellm2018,Sahasrabudhe2020}. Thus, in the temperature regime of about 8~K to 80~K, the very strong field dependence of $\kappa_{ab}$ is straightforwardly interpreted to originate from phonon scattering off such magnons with small momenta \cite{Hentrich2018}. Quite clearly, such magnons are consistent with microwave absorption data \cite{Ponomaryov2017,Wellm2018}. In fact, using a parameter set which results from fits to inelastic neutron scattering spectra,
theoretical results for the magnetic excitation spectrum both in ED and SWT \cite{Winter2018,Wolter2017,Janssen2019}, yield a magnon mode at the $\Gamma$ point which beyond the critical field $B_{c}$ reproduces remarkably well the ESR data and our results for $\hbar\omega_0$ both concerning the mode's energy as well as its linear field dependence.
Since these excitations are the lowest-lying ones near the $\Gamma$ point, the conjectured mode at $\hbar \omega_1$ must be of a different nature. While its sublinear field dependence may be consistent with that of quantum critical excitations at the $M$ point \cite{Wolter2017,Janssen2019}, the mode energy $\hbar\omega_1$ which we extract from our data is at least by a factor of 10 smaller than the magnon gap energy as calculated in SWT and experimentally extracted from specific-heat measurements \cite{Wolter2017,Janssen2019}. This inconsistency appears to rule out the possibility that the low-energy magnetic scattering originates from conventional high-field magnons (or bound states thereof).

Alternatively, it seems quite possible that the scattering arises from defect-induced low-energy excitations. Signatures of such excitations appear in NMR data on nominally clean $\alpha$-RuCl$_3$ \cite{Baek2017}, and they have been studied recently in some detail in Ir-doped $\alpha$-RuCl$_3$ \cite{Baek2020}. The excitations have been discussed in terms of random spin singlets with a broad distribution of (relatively small) binding energies. The thermal excitations of such singlets inevitably scatter phonons and thus one can expect an impact on the phonon heat conductivity. Since the external magnetic field increasingly polarizes such singlets, their density of states is field-dependent, and the phonon scattering should become weaker with increasing field. In this case, the extracted energy scale $\hbar\omega_1$ does not directly correspond to a specific mode energy but rather is an effective quantity which accounts for the field-induced depletion of random singlets. We note that the relatively small scattering intensity of the low-energy excitations, $C_1\ll C_0$, is fully compatible with them being defect-induced.
A detailed modelling of phonon scattering by random singlets requires more insights into the microscopics of disordered Kitaev magnets at elevated fields and is left for future work.

In summary, our high-field/low-temperature study of the phonon heat conductivity of the Kitaev material $\alpha$-RuCl$_3$ reveals that novel low-energy spin excitations exist which are incompatible with conventional magnon-like excitations. A possible origin for these low-energy modes are random singlets which could emerge from natural impurities in the system. Thus, our study underpins the intriguingly rich quantum nature of the ground state $\alpha$-RuCl$_3$ and calls for further experimental and theoretical studies.

\begin{acknowledgments}

This work has been supported by the Deutsche Forschungsgemeinschaft through SFB 1143 (project-id 247310070), the W\"urzburg-Dresden Cluster of Excellence on Complexity and Topology in Quantum Matter -- \textit{ct.qmat} (EXC 2147, project-id 390858490), the Emmy Noether program (JA2306/4-1, project-id 411750675), and through the projects HE3439/12 and HE3439/13. This work has further been supported by HFML-RU/NWO-I, member of the European Magnetic Field Laboratory (EMFL) as well as the
European Research Council (ERC) under the European
Union's Horizon 2020 research and innovation programme
(Grant Agreement No. 647276-MARS-ERC-2014-CoG).
\end{acknowledgments}



\bibliography{rucl,frustrated_transport}

\section{Appendix}

\subsection{Fitting of $\kappa_{ab}$ data}

The temperature-dependent data were fit via an analogous routine as described in very detail in the supplement of \cite{Hentrich2018}.

More specifically, we use the Callaway model \cite{Callaway1959,Callaway1961} for analysing the phononic heat conductivity. The model yields the low-$T$ approximation
\begin{equation}
 \kappa_{ab}  (T) = \frac{k_B}{2\pi^2v_s}\left(\frac{k_BT}{\hbar}\right)^3 \int_0^{\Theta_D/T}\,\,
\frac{x^4e^x}{\left(e^x-1\right)^2}\tau_c(\omega,T)\ dx, \label{Callaway}
\end{equation}
with  Boltzmann's constant $k_B$, Planck's constant $\hbar$, the Debye temperature $\Theta _\mathrm{D}$, $x = \hbar\omega / k_BT$, and the effective phonon scattering rate $\tau ^{-1} _c$ which depends on both $\omega$ and $T$.

For conventional phononic systems (i.e. non-magnetic, electrically insulating crystals), the effective phonon scattering rate in Eq.~\eqref{Callaway} is composed of conventional scattering mechanisms, for which empirical expressions are well established, viz. phonon-phonon umklapp scattering $\tau_P^{-1}= A\, T \omega ^3 e^{- \Theta _D/b T}$, phonon-defect scattering $\tau_D^{-1}= D \omega ^4$ and phonon-boundary scattering $\tau_B^{-1}=v_s L^{-1}$:
\begin{equation}
\tau_{c,0}^{-1}=\tau_P^{-1}+\tau_D^{-1}+\tau_B^{-1}. \label{matthiessen}
\end{equation}

The phenomenological magnetic scattering rate $\tau_\mathrm{mag}^{-1}$ adds to the the rate as
\begin{equation}
\tau_c^{-1}= \tau_{c,0}^{-1}+\tau_\mathrm{mag}^{-1}. \label{comb}
\end{equation}

Table~\ref{tab:bindep} lists the field independent phononic parameters of $\tau_c ^{-1}$ as well as the cutoff energy (Kitaev interaction) $K$ used for the fits shown in Fig.~3 of the main article.
Note that data of sample 1 and sample 2, collected at HFML and IFW Dresden, were scaled prior to fitting them to the Callaway model, allowing for treatment as a single data set with common phononic parameters, spanning the entire $T$- and $B$ range.
Table~\ref{tab:bdep} lists the fitting parameters of the magnetic scattering term $\tau_\mathrm{mag}^{-1}=\tau_\mathrm{mag,0}^{-1}+\tau_\mathrm{mag,1}^{-1}$
used for the fits shown in Fig.~3 of the main article.
It is remarkable that a field dependent spin-phonon-coupling $C_0 $ must be allowed for in order to satisfactorily explain the data with our simple model. The so found spin-phonon-coupling increases roughly linearly by about $30~$\% over the investigated field range.

\subsection{Illustration of $\kappa_{ab} (B,T)$ for a single magnetic scattering term}

For the illustrations shown in Fig.~2 (c) of the main article, a single mode model was used with realistic fitting parameters (Tab.~\ref{tab:bindep}). The employed temperatures in the three panels of Fig. 2c) are $T=0.7$~K (bottom panel), $T=16$~K (center panel) and $T=128$~K (top panel).
Values for $\hbar \omega _0(B) / k_B$ were obtained by linearly extrapolating previous results ~\cite{Hentrich2018}, according to $\hbar \omega _0(B) / k_B =  4.02~\mathrm{K}/\mathrm{T} \times B - 22.2$~K.


\onecolumngrid

\vspace{2cm}


\begin{table*}[!hb]

  \centering
  \begin{tabularx}{0.7 \textwidth}{c|c|c|c|c|c}

 \hspace{.1cm} Parameter \hspace{.1cm} &  $ \mspace{8mu} A$ (\SI{10^{-31}}{K^{-1}s^2})$\mspace{8mu}$ &  $\mspace{8mu} b \mspace{8mu}$ &  $\mspace{8mu} D $ (\SI{10^{-43}}{s^3})$\mspace{8mu} $ & $\mspace{8mu} L$ (\SI{10^{-3}}{m}) $\mspace{8mu} $& $\mspace{8mu} K/ k_B $ (\SI{}{K}) $\mspace{8mu} $ \tabularnewline	
\noalign{\smallskip} 	\hline \noalign{\smallskip}
	$\kappa_{ab}$ & \SI{8.27}{}  & $\mspace{20mu} \SI{4.21}{}\mspace{20mu}$ & \SI{5.26}{}  & \SI{0.19}{}& \SI{59.5}{}\tabularnewline \noalign{\smallskip} \hline
	
	\end{tabularx}
	\caption{Field independent parameters for fitting $\kappa_{ab}$ }
  \label{tab:bindep}
\end{table*}

\begin{table*}[!hb]

  \centering

  \begin{tabularx}{0.625\textwidth}{c|c|c|c|c}

{Magnetic Field} $(\SI{}{T})\mspace{3mu}$   			&
$\mspace{8mu}\hbar \omega _0 / k_B (\SI{}{K})\mspace{8mu}$  	&
$\mspace{8mu} C_0 (\SI{10^{9}}{s^{-1}})\mspace{8mu}$ 				&
$\mspace{8mu}\hbar \omega _1 / k_B (\SI{}{K})\mspace{8mu}$ 		&
$\mspace{8mu} C_1 (\SI{10^{7}}{s^{-1}})\mspace{8mu}$ \tabularnewline	
\noalign{\smallskip} 	\hline \noalign{\smallskip}

10 &  0 &  2.58 & 1.16  &  4.5 																				\tabularnewline 
12 &  14.3 &  2.81 & 2.11 &  4.5 																				\tabularnewline 
14 &  26.1 &  2.99 &  3.00 &  4.5 																				\tabularnewline 
16 &  36.0 &  2.9 & 3.49 &  4.5 																				\tabularnewline 
18 &  41.6 &  2.89 & 4.10  &  4.5 																				\tabularnewline 
20 &  50.2 &  2.76 &  3.94 &  4.5 																				\tabularnewline 
22 &  60.5 &  2.74 & 4.28  &  4.5 																				\tabularnewline 
24 &  73.5 &  2.96 & 4.81  &  4.5  																			\tabularnewline 
26 &  85.3 &  2.97 &  5.09 &  4.5  																			\tabularnewline 
28 &  97.8 &  3.21 &  5.46 &  4.5  																			\tabularnewline 
30 &  112.3 &  3.39 &  5.66 &  4.5  																			\tabularnewline 
33 &   &   & 5.94  &  4.5   																			\tabularnewline 
	\end{tabularx}

	\caption{Fitting parameters for the magnetic scattering term (Eq.~\ref{eq:triplet}) for fitting $\kappa_{ab}$.}
  \label{tab:bdep}
\end{table*}


\newpage

\twocolumngrid

\subsection{Fitting of the field dependence of $\hbar \omega_1$}

\begin{figure}[h!]
\centering
\includegraphics[width=12 cm]{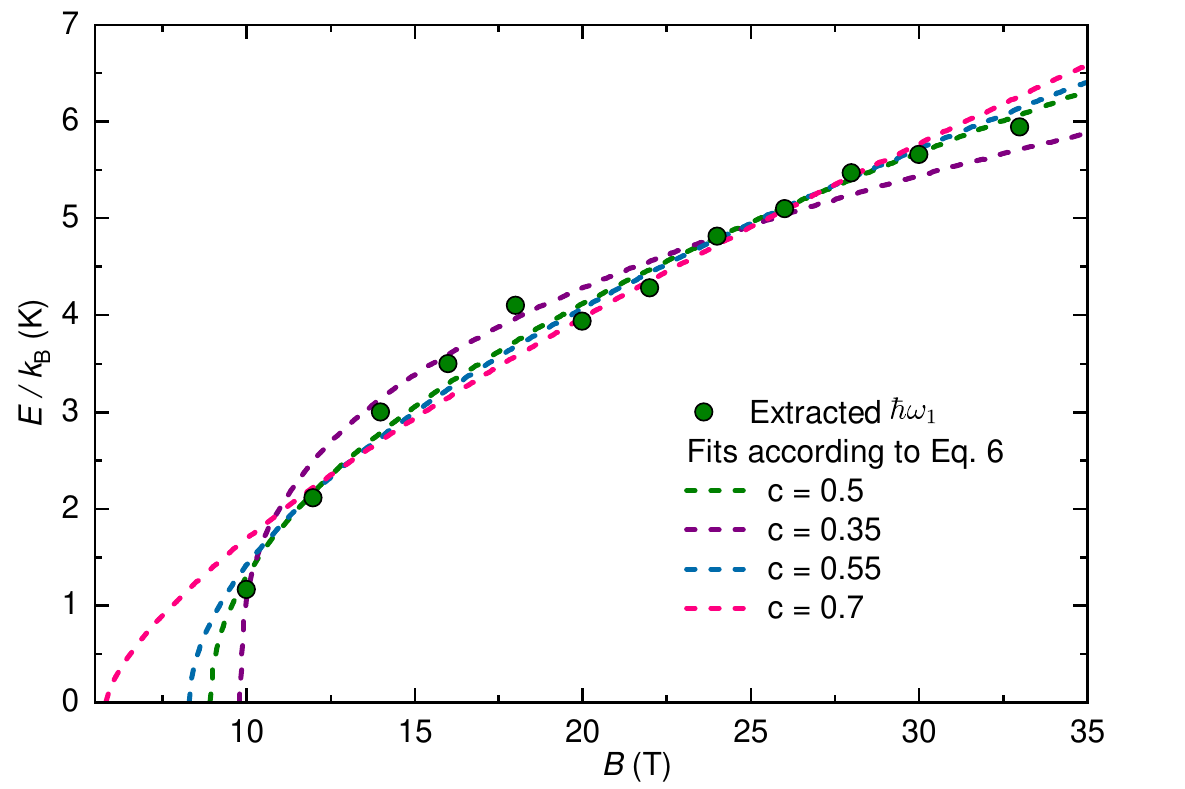}
\caption{Field dependence of  $\hbar \omega _1 (B)$ (green circles) and multiple fits thereof according to Eq.~\ref{eq:powerlaw}.}
\label{fig:exponents}
\end{figure}

Figure~\ref{fig:exponents} shows the obtained field dependent values of $\hbar \omega_1$ in detail. As discussed in the main text, the field dependency of $\hbar \omega_1$ may be described by a power law according to
\begin{equation}
\hbar \omega_1 (B)/k_B = a\left(\left(B-B_c\right)/B_c\right)^c~,
\label{eq:powerlaw}
\end{equation}
where $c=1/2$ leads to good agreement with our data. The corresponding fit, depicted in the inset of Fig.~3~(a) of the main text, is reproduced in Fig.~\ref{fig:exponents} as a green dashed line.
As can be inferred from several additional power law fits plotted in Fig.~\ref{fig:exponents} as dashed lines, mediocre fits can also be obtained for the approximate range $0.4 \le c \le 0.55$.
The pink dashed line illustrates the best fit with $c=0.7$ according to the previous analysis of specific-heat data \cite{Wolter2017}, where a rather unrealistic $B_c \approx 5.8$~T is used.

All parameters of the fits depicted are listed in table~\ref{tab:paras_abc}.


\begin{table}[!tbh]

  \centering
  \begin{tabularx}{ .7\columnwidth}{c|c|c|c}

 \hspace{.1cm} Fit \# \hspace{.1cm} &  $ \mspace{8mu} a$ (\SI{}{K})$\mspace{8mu}$ &  $\mspace{12mu} B_c(\SI{}{\tesla}) \mspace{12mu}$ &  $\mspace{20mu} c \mspace{20mu} $ \tabularnewline	
\noalign{\smallskip} 	\hline \noalign{\smallskip}
	1 & 3.68 & 8.91 & 0.5 \tabularnewline
	2 & 3.49 & 9.81 & 0.35 \tabularnewline
	3 & 3.41 & 8.29 & 0.55 \tabularnewline
	4 & 1.77 & 5.82 & 0.7 \tabularnewline \noalign{\smallskip}
	
	\end{tabularx}
	\caption{Parameters of power law fits for $\hbar \omega_1 (B)/k_B$, as depicted in Fig.~\ref{fig:exponents}. }
  \label{tab:paras_abc}
\end{table}


\onecolumngrid
\vspace{1cm}

\end{document}